# Potential improvement for elastocaloric performances of shape memory alloys through comparing with natural rubber

Zhongjian Xie[1], Gael Sebald, Daniel Guyomar

Université de Lyon, INSA-Lyon, LGEF EA682, 8 rue de la Physique, 69621 Villeurbanne, France

**Abstract**

The elastocaloric (eC) effects of natural rubber (NR) and shape memory alloys (SMAs) are compared and potential improvement for eC performances of SMAs is proposed. Both the plateau and hysteresis of stress-strain curve in NR and SMAs are observed. They are attributed to the similar phase transformation: strain-induced crystallization (SIC) in NR and martensitic transformation in SMAs. For NR, the drawback of large strain can be reduced by pre-strain. For SMAs, the stress change with strain is smaller in stress plateau regime. Thus, the drawback of large dynamic stress of SMAs is proposed to be reduced by a static pre-stress for working directly in this stress plateau regime. Choosing an appropriate pre-strain (pre-stress) and strain amplitude can make eC effect of NR working in the phase transformation (SIC) regime. This method can improve both the eC strength and fatigue life in NR. Due to the similar phase transformation (martensitic transformation) of SMAs to NR, this method is also proposed to improve the eC performances of SMAs.

The solid materials can undergo an isothermal entropy change or an adiabatic temperature change upon the application or withdrawal of uniaxial stress, which is the elastocaloric (eC) effect. The eC effect of shape memory alloys (SMAs) has attracted more and more attention due to the large adiabatic temperature change. It is an order of magnitude higher than the magnetocaloric (MC) effect.[1–5] The refrigerator systems similar to active magnetic regenerator (AMR) by using eC effect of SMAs have been proposed by Tusek *et al.*[5] and Qian *et al.*[6] A real set-up has been built by Schmidt et al.[7] Previous studies suggested that Ni-Ti alloy had adiabatic temperature changes of 17-23 K,[3,8] whereas the Cu-based alloys had adiabatic temperature changes of 11-19 K.[5,9] Fe-Rh[10,11] and Eu-Ni-Si-Ge[12] alloys had adiabatic temperature change of 8.7 K and 14 K, respectively. Despite the large temperature change of SMAs, their application is still limited by the large stress and short fatigue life. Although a 10 million cycles has been demonstrated in Ti-rich $Ti_{54}Ni_{34}Cu_{12}$ by Chluba *et al.*,[13] it was tested at temperature of 70 $^{o}$C with austenite-finish temperature of 65 $^{o}$C, which limits its near room temperature cooling application. Thus, the material combining both a long fatigue life and appropriate transition temperature need to be further investigated.

In fact, the eC effect of NR was the earliest known one, which was observed by Gough and further investigated by Joule (also called the Gough-Joule effect).[14] NR shows a large eC temperature change of about 10 $^{o}$C at strain of 5.[15–17] However, the eC effect of NR is only considered for cooling application recently.[18–22] Its eC strength can be improved by choosing an appropriate minimum strain, which can be chosen for pre-strain.[19] Its indirect measurement was compared with direct measurement.[20] Most importantly, the fatigue property of eC effect of NR was tested. No crack occurred and a stable eC temperature change was observed after $1.7 \times 10^5$ cycles (in press). For elasticity research of NR, it has existed for a long time.[23–25] A comprehensive study of dynamic fatigue life of NR was reported in 1940.[26] It can facilitate the eC research in NR and may be further analogized to the eC study in SMAs.

---
[1] zhongjian.xie521@gmail.com
  gael.sebald@insa-lyon.fr

In this letter, the eC effects of NR and SMAs are compared. The similar phase transformation in NR and SMAs responsible for their eC behavior are shown. A way to reduce the large stress of SMAs is proposed. Moreover, the method of choosing an appropriate pre-strain and strain amplitude to improve the eC performances of SMAs is proposed similar to the NR.

The stress-strain behaviors of one kind of SMAs, NiTi,[27] and NR[28] are compared. Some similar phenomena can be observed from Fig. 1. The stress curves for both SMAs and NR can be divided into three regimes: an elastic behavior at small strains, a plateau at middle strains and a stress upturn at large strains. The similar phase transformation occur in both SMAs and NR: the martensitic transformation in SMAs[2] and strain-induced crystallization (SIC) in NR,[19] respectively. Martensitic transformation is a first-order diffusionless structural transformation from cubic austenite phase to a lower symmetry martensitic phase.[29] The stress plateau of SMAs is due to the occurrence of stress-induced martensitic transformation while the stress upturn is due to the further deformation in martensitic phase.[27] SIC is the transformation of the network chains from amorphous phase to crystalline phase upon the application of the stress. In NR, SIC occurs at around strain of 3[28,30–32] and it consists of two stress effects: stress relaxation[33,34] and stress hardening.[35] The partial crystalline chain can relax the remaining amorphous chain and decrease the stress. This is referred as stress relaxation effect of SIC. The crystallite acts as crosslink and binds plenty of chains.[36] Thus, the network chain density will increase, resulting in the stress increase. This is referred as stress hardening effect of SIC. The stress plateau of NR is due to the stress relaxation effect while the stress upturn is due to the stress hardening effect of SIC. In conclusion, for both stress-strain behaviors of SMAs and NR, the elastic behavior at small strains occurs in the original phase, the plateau at middle strains comes from the phase transformation, and the stress upturn comes from the deformation in the new formed phase.

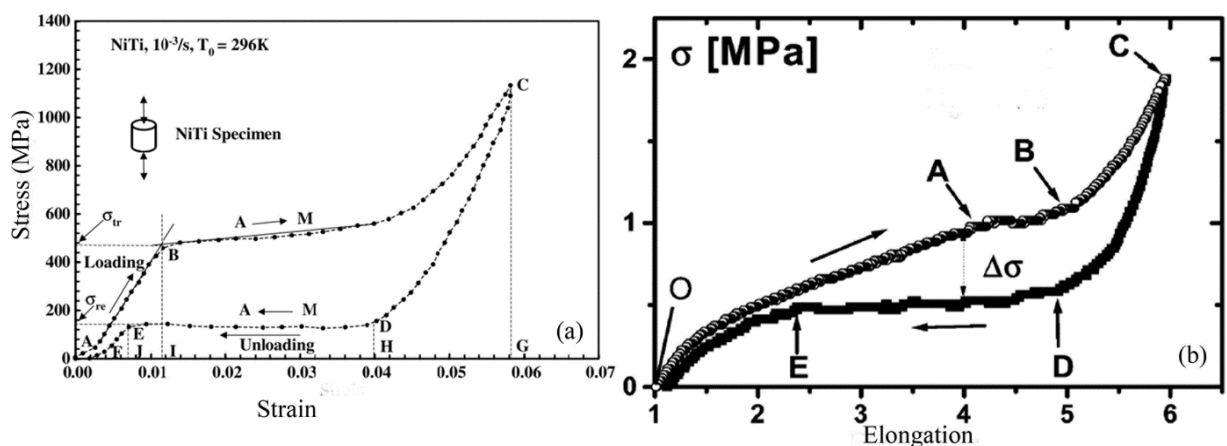

Fig. 1. (a) Stress-strain curve of one kind of SMAs, NiTi[27] *[Reprinted from Mechanics of Materials, 38, S. Nemat-Nasser and W.-G. Guo, Superelastic and cyclic response of NiTi SMA at various strain rates and temperatures, 463-474, Copyright (2006), with permission from Elsevier]* and (b) stress-elongation curve of NR.[28] *[Reprinted (adapted) with permission from Macromolecules, 36, S. Trabelsi, P.A. Albouy, and J. Rault, Crystallization and melting processes in vulcanized stretched natural rubber, 7624-7639. Copyright (2003) American Chemical Society]*

To obtain sufficient cooling capacity, the strain needed for eC material determines the size of the cooling device while the stress needed determines the cooling device is easy or difficult to be performed. Stress and strain are two conjugated variables. The NR is a soft material with Young modulus of several MPa, resulting in a low tensile stress in NR. Accordingly, the NR needs to be stretched several times of its original length. The large deformation is a main drawback for a compact cooling device. To overcome this drawback, a pre-strain is applied in NR.[19] SMAs are hard materials with a small strain (~10%) and large stress (several hundreds of MPa). In Tusek's model,[5] for 500 W of the cooling power, the system requires an applied force of about 180 kN (stress of 900 MPa) for Ni-Ti alloy and about 50 kN (stress of 275 MPa) for Cu-Zn-Al. This large force is difficult to be provided. Moreover, it will make the SMAs sample difficult to be hold in the clamp. A smaller stress change with strain is expected in the stress plateau regime (Fig. 1(b)). Thus, similar to the method of pre-strain for NR, an application of static pre-stress in SMAs to this stress plateau regime may help to decrease the large stress in dynamic stretching and thus to improve the tensile stability in SMAs.

Moreover, both the stress-strain curves in SMAs and NR show the hysteresis.[3,13,20,28] For SMAs, the hysteresis of the stress-strain curve is due to the hysteresis of the martensitic transformation, originating from the irreversible energy dissipation.[37] It results in an increased input work to perform the eC cooling cycle.[3] For NR, the hysteresis is originated from the interaction of the stress relaxation and stress hardening effects of SIC. In the loading process, the stress hardening effect of SIC is dominant, i.e. SIC mainly increases the stress. In the unloading process, the stress relaxation effect of SIC is dominant. For a given strain, the crystallinity during the unloading process is higher than the corresponding one in the loading process.[32] Thus, the stress during the unloading process is lower than the corresponding one in the loading process due to the dominate stress relaxation effect of SIC.

Similar to stress-strain behaviors of SMAs and NR, their adiabatic temperature changes in function of strain also show the hysteresis. Both the adiabatic temperature change of SMAs and NR originates from the latent heat of phase transformation.[38,39] In Fig.2 (a), for Ni-Ti alloy, there is an irreversibility between loading and unloading.[3] This may be due to the occurrence of temporary residual strain, which decreases the temperature change during the unloading.[3] Contrary to Ni-Ti alloy, the temperature increase during loading is smaller than the temperature decrease during unloading for NR, as shown in Fig. 2(b).[19] SIC contributes to the temperature increase and melting contributes to the temperature decrease. The hysteresis in adiabatic temperature change-stain curve of NR is due to the different kinetic properties of SIC and melting.[28,40,41] SIC requires time to be completed while the melting is much faster than SIC.[28] The SIC time is longer than the adiabatic limit (the characteristic time of heat exchange with outer medium). Consequently, part of the temperature increase from SIC can't be measured due to the heat loss with outer medium. Before the unloading process, the crystallinity is larger than that contributes to the temperature increase in loading process. In the unloading process, nearly all the SIC melts and contributes to the temperature decrease due to the faster melting. Thus, the temperature decrease during unloading is larger than temperature increase during loading, resulting in the hysteresis.

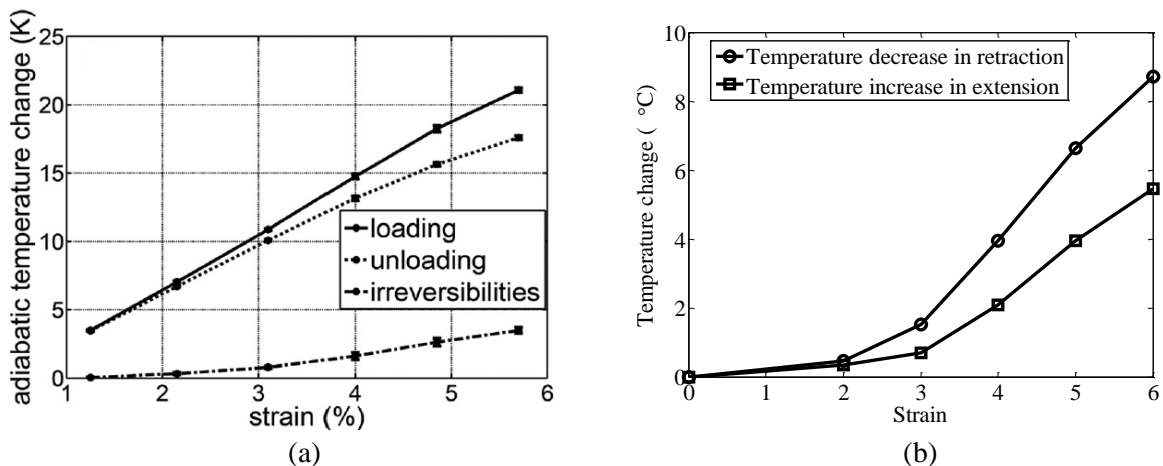

(a) (b)

Fig.2. (a) The adiabatic temperatures change of Ni-Ti wire at different applied strains in loading, unloading process and the irreversibility.[3] *[Reprinted with permission from Journal of Applied Physics, 117, J. Tušek, K.*

The eC strength of NR and SMAs at different strain/stress regime is considered. In NR, a coefficient defined as the entropy change per elongation change ($-\partial s/\partial \lambda$) has been used to describe the eC strength.[19] The entropy change mainly comes from the latent heat of SIC.[39] Choosing pre-strain right before the onset of SIC and a strain amplitude in the SIC onset strain regime, is proved to be optimum for NR to achieve a higher eC strength.[19] It may also be true for SMAs. For SMAs, the eC strength is commonly determined by the temperature change per unit stress change ($\Delta T/\Delta \sigma$).[42] Similar to NR, the temperature change $\Delta T$ of SMAs is mainly from the latent heat of martensitic phase transformation as mentioned previously.[3] Thus, in the stress plateau regime with the existence of martensitic phase transformation, a larger temperature change is expected. Moreover, the stress plateau regime indicates a smaller stress change with strain (Fig. 1(a)). Thus, a higher eC strength of SMAs represented by $\Delta T/\Delta \sigma$ may be achieved in the stress plateau regime. Similar to that in NR, a higher eC strength can be obtained through choosing appropriate pre-strain and strain amplitude. Further experimental work need to be performed on SMAs.

Another important property of the eC material is the fatigue life. Though the eC study of NR is started later than SMAs, its elasticity study has existed for a long time.[24] The NR shows an excellent fatigue performance due to the crack growth resistance of SIC.[43] It is found that, the decreasing strain amplitude can increase the fatigue life of NR (Fig. 3(a)).[26,43] As the strain amplitude decreases from 350% to 25%, the dynamic fatigue life can be increased from $10^3$ to $10^9$ cycles. These fatigue tests are conducted in air. Considering the real cooling device, the eC material should be immersed in some heat transfer fluid, for example, the water. Without the accelerating ageing of NR in air from the oxidation,[44] the fatigue life in water would be longer. The similar fatigue property of SMAs (for example, $Ni_{50}Ti_{40}Cu_{10}$) with NR has been found. Its fatigue life also increases as the stress amplitude decreases (Fig. 3(b)).[45] As the stress amplitude decrease from 200 MPa to 20 MPa, its fatigue life increases from 2000 cycles to 30000 cycles. Other reports on the fatigue test of NiTi show that the fatigue life increases from $10^3$ to $10^7$ cycles as the strain amplitude decreases from the order of 10% (inducing complete martensitic transformation)) to the order of 1%.[46–49] It indicates a dramatic fatigue of SMAs upon full cycle of the stress-induced martensitic transformation. The refrigeration application requires high-cycle loading, so only partial martensitic transformations of SMAs can be used.

Moreover, for the same strain amplitude of NR, different minimum strains can lead to different fatigue lives (Fig. 3(a)). The appropriate minimum strain (pre-strain) of around 200% can result in the longest fatigue life, which may correspond to the SIC onset strain regime. SIC of NR is responsible for the crack growth resistance and thus the good fatigue performance in SIC onset strain regime.[50] The influence of minimum strain on fatigue life of SMAs hasn't been reported. The similar phase transformation may also increase the fatigue life of SMAs similar to NR. Thus, choosing a minimum strain and strain amplitude to make the deformation in the phase transformation onset regime for SMAs may improve their fatigue life. In conclusion, an application of static pre-stress in SMAs to make eC deformation in the phase transformation onset regime directly, which is previously proposed to overcome their drawback of large dynamic stress, may also improve their eC performances (including eC strength and fatigue life).

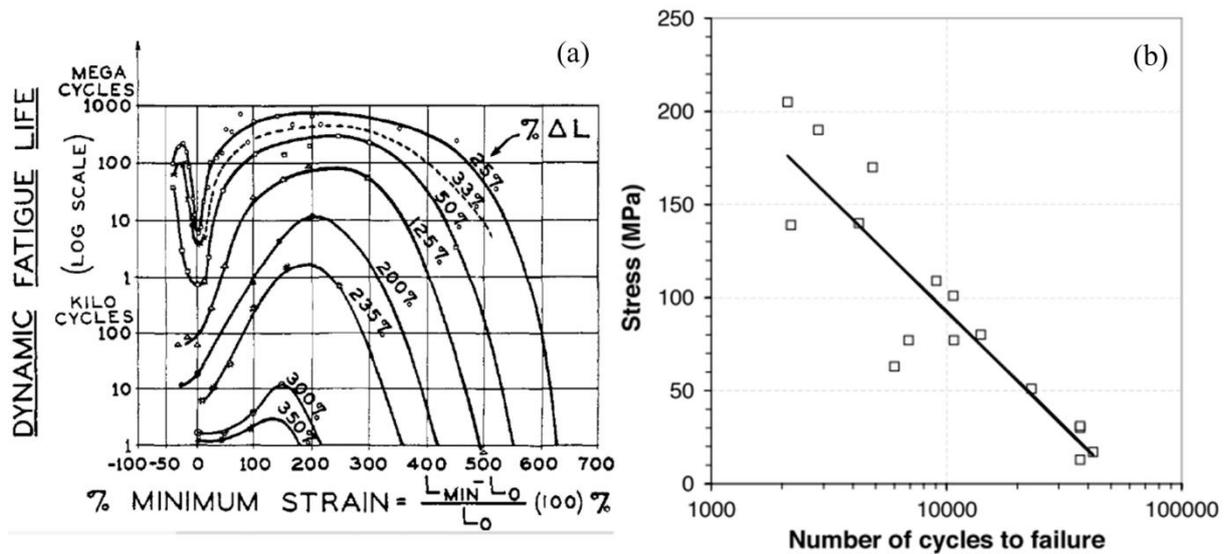

Fig. 3. (a) The fatigue life at different minimum strains and strain amplitudes for NR[26] *[Reprinted with permission from Rubber Chemistry and Technology. Copyright © (1940), Rubber Division, American Chemical Society, Inc.]* and (b) applied stress-fatigue life behavior for $Ni_{50}Ti_{40}Cu_{10}$.[45] *[Reprinted with permission from Smart Materials and Structures, 18, D.C. Lagoudas, D.A. Miller, L. Rong and P.K. Kumar, Thermomechanical fatigue of shape memory alloys, 085021, Copyright [2009], © IOP Publishing]*

In conclusion, for eC effects of NR and SMAs, some similar phenomena are observed, like the plateau and hysteresis of stress-strain behavior, and hysteresis of temperature change-strain behavior. They are all originated from the similar phase transformation: strain-induced crystallization (SIC) in NR and martensitic transformation in SMAs, respectively. The main drawback of NR and SMAs acting as eC materials are their large deformation and large stress, respectively. A pre-strain can be applied to overcome the drawback of large deformation in NR. For SMAs, the deformation for eC effect in the stress plateau regime may need a lower dynamic stress. Thus, a static pre-stress is proposed to be applied, which can make deformation go directly into this stress plateau regime. The temperature change in both NR and SMAs are originated from the latent heat of the phase transformations. Moreover, SIC is responsible for the good fatigue property in NR. Thus, through choosing minimum strain (pre-strain) and strain amplitude, which can make eC effect of NR work in the SIC onset strain regime, both the eC strength and the fatigue life in NR can be improved. This method can also make SMAs work in the martensitic transformation onset regime, which may also lead to higher eC strength and longer fatigue life for SMAs.

In the further work, the reduction of large dynamic stress for SMAs upon a static pre-stress need to be tested. Moreover, the eC strength and fatigue life of SMAs upon an appropriate pre-strain and strain amplitude need to be tested.

**Acknowledgements**
This work was financially supported by China Scholarship Council (CSC).

**References**

[1] X. Moya, S. Kar-Narayan, and N.D. Mathur, Nat. Mater. **13**, 439 (2014).

[2] E. Bonnot, R. Romero, L. Mañosa, E. Vives, and A. Planes, Phys. Rev. Lett. **100**, 125901 (2008).

[3] J. Tušek, K. Engelbrecht, L.P. Mikkelsen, and N. Pryds, J. Appl. Phys. **117**, 124901 (2015).

[4] J. Cui, Y. Wu, J. Muehlbauer, Y. Hwang, R. Radermacher, S. Fackler, M. Wuttig, and I.

Takeuchi, Appl. Phys. Lett. **101**, 073904 (2012).

[5] J. Tušek, K. Engelbrecht, R. Millán-Solsona, L. Mañosa, E. Vives, L.P. Mikkelsen, and N. Pryds, Adv. Energy Mater. **5**, 1500361 (2015).

[6] S. Qian, A. Alabdulkarem, J. Ling, J. Muehlbauer, Y. Hwang, R. Radermacher, and I. Takeuchi, Int. J. Refrig. **57**, 62 (2015).

[7] M. Schmidt, A. Schütze, and S. Seelecke, Int. J. Refrig. **54**, 88 (2015).

[8] J. Cui, Y. Wu, J. Muehlbauer, Y. Hwang, R. Radermacher, S. Fackler, M. Wuttig, and I. Takeuchi, Appl. Phys. Lett. **101**, 25 (2012).

[9] C. Rodriguez and L.C. Brown, Metall. Mater. Trans. A **11**, 147 (1980).

[10] S.A. Nikitin, G. Myalikgulyev, M.P. Annaorazov, A.L. Tyurin, R.W. Myndyev, and S.A. Akopyan, Phys. Lett. A **171**, 234 (1992).

[11] M.P. Annaorazov, S.A. Nikitin, A.L. Tyurin, K.A. Asatryan, and A.K. Dovletov, J. Appl. Phys. **79**, 1689 (1996).

[12] T. Strässle, A. Furrer, Z. Hossain, and C. Geibel, Phys. Rev. B **67**, 54407 (2003).

[13] C. Chluba, W. Ge, R. Lima de Miranda, J. Strobel, L. Kienle, E. Quandt, and M. Wuttig, Sci. **348**, 1004 (2015).

[14] G.A. Holzapfel and J.C. Simo, Comput. Methods Appl. Mech. Eng. **132**, 17 (1996).

[15] L.R.G. Treloar, *The Physics of Rubber Elasticity* (Oxford university press, Oxford, 1975).

[16] S.L. Dart, R.L. Anthony, and E. Guth, Ind. Eng. Chem. **34**, 1340 (1942).

[17] J.C. Mitchell and D.J. Meier, Rubber Chem. Technol. **42**, 1420 (1969).

[18] D. Guyomar, Y. Li, G. Sebald, P.J. Cottinet, B. Ducharne, and J.F. Capsal, Appl. Therm. Eng. **57**, 33 (2013).

[19] Z. Xie, G. Sebald, and D. Guyomar, Appl. Phys. Lett. **107**, 081905 (2015).

[20] Z. Xie, G. Sebald, and D. Guyomar, Appl. Phys. Lett. **108**, 041901 (2016).

[21] Z. Xie, G. Sebald, and D. Guyomar, arXiv Prepr. arXiv1604.04479 (2016).

[22] Z. Xie, G. Sebald, and D. Guyomar, arXiv Prepr. arXiv1604.02686 (2016).

[23] G. Allen, U. Bianchi, and C. Price, Trans. Faraday Soc. **59**, 2493 (1963).

[24] P.H. Mott and C.M. Roland, Macromolecules **29**, 6941 (1996).

[25] W. Mars and A. Fatemi, Int. J. Fatigue **24**, 949 (2002).

[26] S.M. Cadwell, R. a. Merrill, C.M. Sloman, and F.L. Yost, Rubber Chem. Technol. **13**, 304 (1940).

[27] S. Nemat-Nasser and W.-G. Guo, Mech. Mater. **38**, 463 (2006).

[28] S. Trabelsi, P.A. Albouy, and J. Rault, Macromolecules **36**, 7624 (2003).

[29] A. Planes and L. Mañosa, Solid State Phys. **55**, 159 (2001).

[30] N. Candau, R. Laghmach, L. Chazeau, J.-M. Chenal, C. Gauthier, T. Biben, and E. Munch, Macromolecules **47**, 5815 (2014).

[31] M. Tosaka, S. Murakami, S. Poompradub, S. Kohjiya, Y. Ikeda, S. Toki, I. Sics, and B.S. Hsiao, Macromolecules **37**, 3299 (2004).

[32] S. Toki, T. Fujimaki, and M. Okuyama, Polymer (Guildf). **41**, 5423 (2000).

[33] P.A. Albouy, A. Vieyres, R. Pérez-Aparicio, O. Sanséau, and P. Sotta, Polymer (Guildf). **55**, 4022 (2014).

[34] P.J. Flory, J. Chem. Phys. **15**, 397 (1947).

[35] J. Rault, J. Marchal, P. Judeinstein, and P. a. Albouy, Eur. Phys. J. E **21**, 243 (2006).

[36] J. Che, C. Burger, S. Toki, L. Rong, B.S. Hsiao, S. Amnuaypornsri, and J. Sakdapipanich, Macromolecules **46**, 4520 (2013).

[37] Y. Liu, in *Shape Mem. Alloy.* (2010), pp. 361–369.

[38] H. Ossmer, C. Chluba, B. Krevet, E. Quandt, M. Rohde, and M. Kohl, J. Phys. Conf. Ser. **476**, 12138 (2013).

[39] L. Thien-Nga, J. Guilie, and P. Le Tallec, in *Eur. Congr. Comput. Methods Appl. Sci. Eng. (ECCOMAS 2012), Austria* (2012), pp. 10–14.

[40] J.R. Samaca Martinez, J.B. Le Cam, X. Balandraud, E. Toussaint, and J. Caillard, Polymer (Guildf). **54**, 2717 (2013).

[41] K. Brüning, K. Schneider, S. V. Roth, and G. Heinrich, Macromolecules **45**, 7914 (2012).

[42] B. Lu and J. Liu, Sci. Bull. **60**, 1638 (2015).

[43] N. Saintier, G. Cailletaud, and R. Piques, Mater. Sci. Eng. A **528**, 1078 (2011).

[44] P.H. Mott and C.M. Roland, Rubber Chem. Technol. **74**, 79 (2001).

[45] D.C. Lagoudas, D.A. Miller, L. Rong, and P.K. Kumar, Smart Mater. Struct. **18**, 085021 (2009).

[46] K.N. Melton and O. Mercier, Acta Metall. **27**, 137 (1979).

[47] G. Eggeler, E. Hornbogen, A. Yawny, A. Heckmann, and M. Wagner, Mater. Sci. Eng. A **378**, 24 (2004).

[48] H. Tobushi, T. Hachisuka, S. Yamada, and P.-H. Lin, Mech. Mater. **26**, 35 (1997).

[49] H. Tobushi, T. Hachisuka, T. Hashimoto, and S. Yamada, J. Eng. Mater. Technol. **120**, 64 (1998).

[50] S. Trabelsi, P.A. Albouy, and J. Rault, Macromolecules **35**, 10054 (2002).